\documentclass[10pt]{iopart}

\usepackage{graphicx}
\newcommand{\ket}[1]{| #1 \rangle}
\newcommand{\bra}[1]{\langle #1 |}

\begin{document}

\title{Accessibility of physical states and non-uniqueness of entanglement
measure}

\author{Fumiaki Morikoshi$^{1,2}$
\footnote{E-mail: fumiaki@will.brl.ntt.co.jp},
Marcelo Fran\c{c}a Santos$^2$ \footnote{E-mail: m.santos@ic.ac.uk} and\\
Vlatko Vedral$^2$\footnote{E-mail: v.vedral@ic.ac.uk}}

\address{$^1$ NTT Basic Research Laboratories, NTT Corporation,\\ 3-1
  Morinosato-Wakamiya, Atsugi, Kanagawa, 243-0198, Japan}
  
\address{$^2$ Optics Section, The Blackett Laboratory, Imperial College \\
   Prince Consort Road, London, SW7 2BW, United Kingdom}

\begin{abstract}

Ordering physical states is the key to quantifying some physical property of
the states uniquely.
Bipartite pure entangled states are totally ordered under local operations
and classical communication (LOCC) in the asymptotic limit and uniquely
quantified by the well-known entropy of entanglement.
However, we show that mixed entangled states are partially ordered
under LOCC even in the asymptotic limit.
Therefore, non-uniqueness of entanglement measure is understood on the basis of
an {\it operational} notion of asymptotic convertibility.

\end{abstract}


\maketitle

\section{Introduction}

Accessibility between two physical states by some physical process is crucial
in being able to compare the states quantitatively.
When there exists an {\it operation} that converts one state into another, we
can derive an ordering between the two states from the accessibility based on
this operation.
This ordering (together with a few other natural assumptions) makes it possible
to define a quantity that compares the states.
However, if it is impossible to convert one state into another in either
direction within a given framework, there exists no coherent way to compare
those two states.

Uniqueness of a measure that quantifies a certain physical property is strongly
related to ordering of states.
When all elements in a given set of physical states can be completely ordered,
i.e. arbitrary two states can be ordered ({\it total order}), we can make at
least one consistent measure that quantifies the set.
However, if there exists no ordering that works globally, i.e. a certain pair
of states cannot be ordered ({\it partial order}), then we fail to find a
consistent way to `align' all the states.
In other words, total order is a necessary condition for a set to be quantified
by the unique measure.

One of the most familiar examples in physics that contain partial order is in
special theory of relativity.
A pair of events in the spacetime that include each other in their
light cone (i.e. the interval between the two events is time-like) are
accessible because one can affect the other by sending some signal.
However, if one is outside of the light cone of the other (i.e. the interval
between the two events is space-like), then it is impossible to connect them
by any physical operation.
Therefore, there exists no unique way of ordering two such states;
different orderings are possible by choosing different reference frames.
Therefore, the set of events is a partially ordered one, which leads to
the well-known non-uniqueness of simultaneity that follows from the
principles of special theory of relativity 
(See Chapter 17 of \cite{Feynman}, for example).
Furthermore, in a modern approach to relativity, a fundamental structure of
spacetime is modeled as a partially ordered set called a causal set 
\cite{Sorkin1987}.

An important example of partial order in a purely mathematical context is the
theory of majorization \cite{Bhatia}, which is a powerful tool for comparing
two vectors and deriving various inequalities between operators.
For example, majorization brings partial order to probability vectors
and leads to useful inequalities of quantities related to entropy in
statistical mechanics \cite{Wehrl1978}.
An intimate relation between majorization and quantum information theory has
also been discovered recently, which we will mention below.

The most beautiful and successful application of the theory of ordering
physical states is in thermodynamics, where all equilibrium states are totally
ordered under adiabatic processes and quantified by the unique measure of
entropy.
Given two equilibrium states (A and B), entropy $S$ distinguishes between
possible and impossible directions of adiabatic processes between the two
states;
A can access B via an adiabatic process iff $S({\rm A}) \le S({\rm B})$.
(If the equality holds, B can also access A and so the process becomes
reversible.)
The uniqueness of entropy was proved by Giles with his axiomatic approach,
which was developed to clarify the structure of thermodynamics \cite{Giles}.
Giles's work is a culmination of the movement towards a more lucid
understanding of the second law of thermodynamics, starting with
Carath\'eodory (See reference~\cite{Landsberg1956}).
This approach has recently been revisited by Lieb and Yngvason
\cite{Lieb1999,Lieb2000}.

It has been shown recently that thermodynamics and theory of pure-state
entanglement share the same mathematical structure, Giles's rigorous set of
mathematical axioms.
Adiabatic processes in thermodynamics correspond to manipulations of bipartite
pure entangled states by local operations and classical communication(LOCC) in
the context of quantum information theory \cite{Vedral2002-2}. 
Therefore, bipartite pure entangled states are totally ordered under LOCC in
the asymptotic limit, and entropy gives the unique measure in this context as
well (known as the von Neumann entropy of entanglement
\cite{Bennett1996-1,Popescu1997}).

In quantum information theory \cite{Vedral2002-1}, quantum entanglement has
been a subject of intensive research because it is a new resource in physics
as well as an indispensable resource in quantum information processing.
As in the case of other physical resources, it is desirable to find a
unique measure of entanglement in order to exploit it effectively and
efficiently. (For review of entanglement measures see
reference~\cite{Horodecki2001}.)
Contrary to the case of bipartite pure states, the unique measure of
entanglement in mixed states has not yet been established.
It has been proved that if two entanglement measures coincide in pure states
but differ in mixed states, then they impose different orderings
\cite{Virmani2000}.
In fact, some entanglement measures proposed so far are different, and it is
commonly believed that we need different entanglement measures depending on
scenarios.

In this paper, we show that mixed entangled states are partially ordered
under an {\it operational} notion of asymptotic LOCC convertibility by using
the monotonicity of entanglement cost and the fact that
positive-partial-transposition (PPT) bound entangled states cannot be converted
into negative-partial-transposition (NPT) entangled states by LOCC.
Thus, we point out that the partial order structure underlies the
non-uniqueness of entanglement measure.
This immediately reveals the reason why Giles's axiom fails in mixed-state
entanglement, especially axiom 5, which reads
{\it
If two states A and B are both accessible from another state C, then A and B
are accessible in either direction (or both).}
This is exactly what distinguishes total order from partial one.
We show the partially ordered structure of mixed entangled states by giving a
counterexample to axiom 5.
(For other natural axioms and details of Giles's approach, see
references~\cite{Giles,Vedral2002-2}.)

The violation of axiom 5 can be seen, for example, in the theory of relativity
mentioned above:
even if two events A and B are accessible from another event C, the events A
and B are not necessarily accessible from each other because one can be
outside of the light cone of the other.
Another example violating axiom 5 is in entanglement manipulation of
bipartite pure states in {\it finite} regimes.
For example, although both of
$\ket{\phi_1}=\frac{1}{\sqrt{2}}(\ket{00}+\ket{11})$ and
$\ket{\phi_2}=\sqrt{\frac{2}{3}}\ket{00}+\frac{1}{\sqrt{6}}\ket{11}
+\frac{1}{\sqrt{6}}\ket{22}$
can be obtained from a maximally entangled state
$\ket{\Phi_3}=\frac{1}{\sqrt{3}}(\ket{00}+\ket{11}+\ket{22})$ by LOCC,
$\ket{\phi_1}$ and $\ket{\phi_2}$ cannot be converted into each other by LOCC
with certainty.
This is a direct consequence of Nielsen's theorem, which connects
entanglement manipulation and majorization mentioned above \cite{Nielsen1999}.
In the following, we will say two states A and B are {\it incomparable} if
they are not accessible from each other.

The rest of this paper is organised as follows.
First we present a counterexample,
and thus prove the partial order structure of mixed entangled states.
Then, we discuss one possible way of recovering total order.
Finally, we conclude the paper with future directions.

\section{Partial order on mixed entangled states}

First, let us rigorously define the accessibility in axiom 5 within our
context.
The asymptotic convertibility under LOCC is defined as follows:
a state $\rho$ is convertible into a state $\sigma$ if and only if for every
(arbitrarily small) real number $\epsilon$, there exists an integer $n_0$, and
a sequence of LOCC $L_n$ such that for any integer $n \ge n_0$ we have
\begin{equation}
 \parallel L_n (\rho^{\otimes n}) - \sigma^{\otimes n} \parallel \le
 \epsilon,
\end{equation}
where $\rho^{\otimes n} = \rho \otimes \rho \cdots \otimes \rho$ represents a
tensor product of $n$ copies of the state $\rho$ and $|| \cdots ||$ denotes the
usual trace norm distance between two mixed quantum states.
Loosely speaking, this means that one state can be converted into another if a
certain number of copies of the former can arrive at an arbitrarily good
approximation of the {\it same} number of copies of the latter via LOCC in the
asymptotic limit.
We will prove that the set of mixed entangled states is a partially ordered one
under this definition of accessibility.
This is our central result in this paper.
While partial order in bipartite pure states in finite regimes
can be turned into total order in the asymptotic limit under this
accessibility, mixed entangled states still retain partial order structure
even in the asymptotic limit.
Note that we consider the convertibility between the {\it same} number of
copies here unlike ordinary argument of transformations between different
numbers of copies, e.g. from $n$ copies to $m$ copies.
In the transformations between different numbers of copies, it is no wonder
that the transformation is possible at least in one direction if sufficiently
many copies of the initial state are prepared.
Thus, the partial order structure does not clearly appear in the framework,
while it can be easily seen in transformations with the same number of
copies as we will show below.
It is enough to consider the convertibility between the same number
of copies to see whether it is possible or not to compare them because
if one state is `larger' than the other, the former should be converted into
the latter even in this framework.

Intuitively, bipartite mixed states that are most likely to fail axiom 5 are
bound entangled states \cite{Horodecki1998}.
Since bound entangled states are mixed states from which no entangled pure
state can be distilled, if we take one of those, $\rho_{\rm AB}$, and a pure
entangled state, $\sigma_{\rm AB}$,  as a pair of possible candidates for a
counterexample, the first half of the proof has already been accomplished by
definition, i.e. $\rho_{\rm AB} \rightarrow \sigma_{\rm AB}$ is
impossible for all pure states $\sigma_{\rm AB}$.
So, all we have to do is to disprove the convertibility in the opposite
direction.

In order to prove that, we take a particular bound entangled state constructed
from an unextendible product basis (UPB) \cite{Bennett1999}.
Suppose both Alice and Bob have three-level quantum systems (qutrits).
Consider the following incomplete orthonormal product basis:
\begin{equation}
 \begin{array}{rcrcl}
 \ket{\psi_1} &=& \ket{0} &\otimes& \frac{1}{\sqrt{2}} (\ket{0}+\ket{1}) \\
 \ket{\psi_2} &=& \frac{1}{\sqrt{2}}(\ket{0}+\ket{1}) &\otimes& \ket{2} \\
 \ket{\psi_3} &=& \ket{2} &\otimes& \frac{1}{\sqrt{2}}(\ket{1}+\ket{2}) \\
 \ket{\psi_4} &=& (\ket{1}+\ket{2}) &\otimes& \ket{0} \\
 \ket{\psi_5} &=& \frac{1}{\sqrt{3}}(\ket{0}-\ket{1}+\ket{2}) &\otimes&
                  \frac{1}{\sqrt{3}}(\ket{0}-\ket{1}+\ket{2}).
 \end{array}
 \label{UPB}
\end{equation}
This incomplete orthogonal basis form a UPB, which means that there exists no
product state orthogonal to all of the above five states.
Consequently, the four-dimensional subspace complementary to this
five-dimensional one does not contain any product states.
Therefore, with a normalization factor, the projection operator onto this
complementary space
\begin{equation}
 \rho_{\rm AB} = \frac{1}{4} \left( I - \sum_{i=1}^{5} \ket{\psi_i}
 \bra{\psi_i} \right)
 \label{Bound}
\end{equation}
turns out to be an entangled state.
It can also be easily seen that this state satisfies the
positive-partial-transposition (PPT) condition because the identity operator
and projections onto product states like $\ket{\psi_i}\bra{\psi_i}$ remain
positive after partial transposition.
Thus, $\rho_{\rm AB}$ is proved to be a PPT bound entangled state.

The important fact about the state $\rho_{\rm AB}$ is that its entanglement
cost $E_{C}(\rho_{\rm AB})$ is positive \cite{Vidal2001-1}, which is
defined as 
$E_{C}(\rho) \equiv \lim_{n \to \infty} E_{f}(\rho^{\otimes n})/n$
\cite{Hayden2001},
where $E_{f}(\rho)$ represents the entanglement of formation of $\rho$
\cite{Bennett1996-2}.
(It is obvious that some amount of entanglement is necessary to construct one
copy of a bound entangled state.
Until quite recently, however, it was an open question whether the entanglement
cost of bound entangled states is also positive \cite{Vidal2001-1}.)
Owing to this property, one can choose a pure entangled state
$\sigma_{\rm AB}= \ket{\phi} \bra{\phi}$ such that
\begin{equation}
 0<E_{C}(\sigma_{\rm AB})<E_{C}(\rho_{\rm AB}).
 \label{def of sigma}
\end{equation}
For simplicity, we choose $\ket{\phi}$ to be an entangled states with
Schmidt number two or three, i.e. a 2$\times$2 or 3$\times$3 system.
Since the entanglement cost $E_{C}$ is an entanglement monotone, i.e. it
cannot increase under LOCC, $\sigma_{\rm AB}$ can never be converted into
$\rho_{\rm AB}$ even asymptotically, i.e.
$\sigma_{\rm AB} \rightarrow \rho_{\rm AB}$ is impossible.
The monotonicity of entanglement cost $E_{C}$ can be easily derived from
the fact that entanglement of formation $E_{F}$ is also an entanglement
monotone.
Note that the above incomparability holds in the sense of the {\it same}
number of copies.
Otherwise, a sufficiently many copies of $\sigma_{\rm AB}$ can always produce
a much smaller number of copies of $\rho_{\rm AB}$ with certainty.

Besides the above fact, note that a maximally entangled state
$\ket{\Phi_3}_{\rm AB}$
can access both $\rho_{\rm AB}$ and $\sigma_{\rm AB}$.
(Although the state $\ket{\Phi_3}_{AB}$ might not be the most efficient
example to produce them, this does not matter here.
One can make one copy of either $\rho_{\rm AB}$ or $\sigma_{\rm AB}$ from
one copy of $\ket{\Phi_3}_{\rm AB}$ by LOCC.)
Therefore, we found a counterexample that two states $\rho_{\rm AB}$ and
$\sigma_{\rm AB}$ are {\it not} convertible into each other in spite of
the fact that both of them can be accessed from the same state
$\ket{\Phi_3}_{\rm AB}$.
This clearly shows that entangled mixed states are partially ordered under LOCC
in the asymptotic limit and thus violate axiom 5.
(We note that although, in Giles's axioms, transformations assisted by
asymptotically negligible amount of auxiliary states are considered, the
undistillable property of bound entanglement remains unchanged even with an
assistance of auxiliary entangled states \cite{Vidal2001-1}.)

\begin{figure}[htb]
 \begin{center}
  \includegraphics{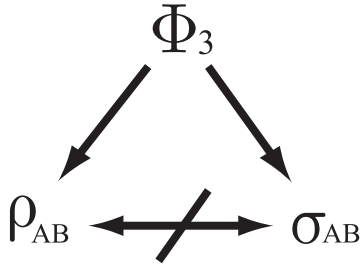}
 \end{center}
  \caption{\label{Fig1} Partial order structure in mixed entangled states.
 $\ket{\Phi_3}=\frac{1}{\sqrt{3}}(\ket{00}+\ket{11}+\ket{22})$,
 $\rho_{AB}$ is a bound entangled state defined in equation~(\ref{Bound}), and
 $\sigma_{AB}$ is a pure entangled state satisfying
 $0 < E_C (\sigma_{AB}) < E_C (\rho_{AB})$.
 The pure state $\sigma_{AB}$ can be replaced with any NPT entangled state
 $\chi_{AB}$ satisfying the same condition.
 }
 \end{figure}

In the above argument, we chose a pure state as $\sigma_{\rm AB}$ for
simplicity.
However, it can be replaced with distillable mixed states $\chi_{\rm AB}$ such
that $0<E_{C}(\chi_{\rm AB})<E_{C}(\rho_{\rm AB})$ because PPT bound entangled
states cannot be converted into NPT ones by LOCC, i.e.
$\rho_{\rm AB} \rightarrow \chi_{\rm AB}$ is impossible for all NPT states
$\chi_{\rm AB}$.
Thus, the above also holds for any such $\chi_{\rm AB}$.
(Furthermore, the monotonicity of other entanglement measures can also be used
for the above argument instead of entanglement cost.)
Generally, it is concluded that any PPT bound entangled state with positive
entanglement cost always has incomparable states in the NPT regime.
(See figure~\ref{Fig1}.) 
These states are the counterexamples to axiom 5, which prevents us from
applying Giles's approach to mixed entangled states within the framework of
LOCC convertibility.

Therefore, we have proved that the set of mixed entangled states is a partially
ordered one under the operational notion of asymptotic convertibility with
LOCC, which underlies the non-uniqueness of entanglement measure.
Since there is no {\it operational} way to link incomparable states, there
exists no way of assigning meaningful amounts of entanglement to them that
could determine which state is more entangled.
In other words, it is the partial order that allows us to use various
entanglement measure known so far without any contradiction.
Metaphorically speaking, we have shown a sort of `relativity' of entanglement
measure under asymptotic convertibility with LOCC, which means that there
exists no {\it absolute} entanglement measure at least under LOCC.

\section{Recovering total order}

Next we discuss a possibility of recovering total order from partial order,
which might lead to the unique measure of entanglement.
Obviously, an extra operation besides LOCC will be necessary to achieve it.
One naive way is restoring quantum information discarded into the environment,
which can be seen as follows.

First, let us think about how the total order structure of pure entangled
states changes into partial one during the formation process of mixed
entangled states.
Imagine a process of making the state $\rho_{\rm AB}$ in equation~(\ref{Bound})
from a maximally entangled state by LOCC.
The state can be written as $\rho_{\rm AB}=\frac{1}{4} \sum_{i=6}^9
\ket{\psi_i} \bra{\psi_i}$, where $\ket{\psi_i} \ (i=6,\ldots,9)$ is an
entangled basis complementary to the UPB in equation~(\ref{UPB}).
One way of making $\rho_{\rm AB}$ is as follows
(it is not necessarily the most efficient way from the viewpoint of the amount
of entanglement invested).
Alice and Bob first prepare a three-level maximally entangled state
$\ket{\Phi_3}_{\rm DB}$ between them.
Besides this, Alice prepares two qutrits, A, B$'$, and a four-level ancilla C
(e.g. two qubits) locally, with which she makes a superposed state
$\ket{\omega}_{\rm CAB'} = \frac{1}{2} \sum_{i=6}^{9} \ket{i}_{\rm C}
\ket{\psi_i}_{\rm AB'}$.
By teleporting the system B$'$ to Bob with the previously shared entanglement
$\ket{\Phi_3}_{\rm DB}$, they succeed in constructing a state
$\ket{\psi}_{\rm CAB} = \frac{1}{2} \sum_{i=6}^{9} \ket{i}_{\rm C}
\ket{\psi_i}_{\rm AB}$ between them.
This state changes into the mixed state $\rho_{\rm AB}$ immediately after
Alice throws away the system C into the environment.

Generally, the total order structure survives until Alice discards the
information because the entire state is just a pure state.
Therefore, if she could retrieve the quantum information from the environment,
the total order would be recovered as well.
Although this may appear impractical if we try to find {\it realistic}
operations that make it possible to restore quantum information from the
environment, it is possible in principle.
In fact, we can also imagine some artificial situation, which might be
plausible in the context of information processing, where Alice gives the
system C to the third party, e.g. Charlie, not to the environment.
In this situation, restoring the quantum information is not impractical at all.
In other words, if we purify (mathematically) a bipartite mixed state
$\rho_{\rm AB}$ into a pure state $\ket{\psi}_{\rm CAB}$ by introducing a local
ancilla C at Alice's side {\it virtually}, then all states become pure states
and we should recover total order.

With this restoring process, we can tell which state of incomparable states
would have more entanglement than the other, if Alice had not lost the quantum
information in the formation process.
We quantify entanglement of pure states by using entropy of entanglement as
usual:
if the restored state is $\ket{\psi}_{\rm CAB}$, then we define the amount of
entanglement $E(\rho_{\rm AB})$ between Alice(CA) and Bob(B) as the von Neumann
entropy of Bob's reduced density matrix
${\rm tr_{CA}} (\ket{\psi}_{\rm CAB} \bra{\psi})$.
It is shown below that the bound entangled state $\rho_{\rm AB}$ would have
more entanglement than the incomparable state
$\sigma_{\rm AB}=\ket{\phi}\bra{\phi}$.
(The approach here is similar to the definition of entanglement of purification
\cite{Terhal2002}, where the entanglement is minimized over all possible
purification in both Alice's and Bob's sides.
Thus, our quantity is always greater than this.)

It is easily seen that
$E(\rho_{\rm AB}^{\otimes n}) \ge E_f(\rho_{\rm AB}^{\otimes n})$
due to the concavity of the von Neumann entropy \cite{Bennett1996-2}.
Thus, we also have
$\lim_{n \to \infty} E(\rho_{\rm AB} ^{\otimes n})/n
= E(\rho_{\rm AB}) \ge E_C(\rho_{\rm AB})
=\lim_{n \to \infty} E_f(\rho_{\rm AB} ^{\otimes n})/n$.
Note that $E$ and $E_C$ coincide for pure states.
Since we chose $\sigma_{\rm AB}$ according to equation~(\ref{def of sigma}),
$E$ of the bound entangled state is always greater than that of the
incomparable pure state, i.e.
$E(\sigma_{\rm AB})=E_C(\sigma_{\rm AB}) < E_C(\rho_{\rm AB})
\le E(\rho_{\rm AB})$.
Therefore, any incomparable states with equation~(\ref{def of sigma}) satisfy
$E (\sigma_{\rm AB}) < E (\rho_{\rm AB})$.

We have considered a simple example of recovering total order by
restoring information from the environment.
Besides the above case, we can conceive other scenarios with different
formation processes.
For example, Alice could perform a measurement on the system C and then
discard the classical information of the outcome instead of discarding
the quantum system directly.
In this case, in order to recover total order, she would restore just
classical information and generally end up in a different pure state from
the original one.
A striking example is $\frac{1}{2} (\ket{00}_{\rm AB} \bra{00}
+ \ket{11}_{\rm AB} \bra{11})$, which will be recovered to a GHZ state in the
former scenario, while it will be recovered to just a separable state in the
latter.
(In fact, the above quantification of $E$ assigns a positive value one to
this state.
This unpleasant quantification is due to the prescription designed only to
compare incomparable states.)
Therefore, it is fair to say that recovered total order is highly dependent
on how information was lost and retrieved.
(See also reference~\cite{Henderson2000} in a different context of
entanglement vs information loss.)

\section{Discussion and conclusion}

It is worth investigating the partial order structure of
mixed entangled states in detail.
Although it was proved only for bound entangled states, we believe the partial
order structure is embedded in distillable states as well.
It is also known that there exist distillable states for which entanglement
cost is strictly greater than distillable entanglement
\cite{Vidal2001-2,Vidal2002}.
Thus, this gap between entanglement cost and distillable entanglement also
leads to the same argument as in the case of bound entangled states just
proved:
suppose $\rho_{\rm AB}$ is a distillable mixed state such that
$E_{D}(\rho_{\rm AB}) < E_{C}(\rho_{\rm AB})$.
Then, $\rho_{\rm AB}$ is incomparable with a pure state whose entropy of
entanglement is sandwiched between $E_{D}(\rho_{\rm AB})$ and
$E_{C}(\rho_{\rm AB})$.
(The monotonicity of distillable entanglement is invoked here instead of
the fact that PPT bound entangled states are not accessible to NPT entangled
states.)
Therefore, it is concluded that the partial order structure lies in
distillable states as well.

In summary, we proved the partial order structure of mixed entangled states
under the operational notion of asymptotic convertibility with LOCC, which
makes it possible to understand the non-uniqueness of entanglement measure 
based on a very general argument on physical accessibility.
The origin of the partial order is likely to be closely related to
information loss.
An important future direction is finding out exactly how the
thermodynamical structure of pure states breaks down when mixedness appears
in entanglement.

\ack

FM appreciates the warmth and hospitality of the quantum information group
at Imperial College London.
MFS acknowledges the support of CNPq.
VV is supported by European Community, Engineering and Physical Sciences
Research Council, Hewlett-Packard and Elsag Spa.

\end{document}